\documentclass[12pt]{article}
 \usepackage{graphicx}
 \usepackage[cp1251]{inputenc}

\begin{document}
 \noindent {\footnotesize\it
   Astronomy Letters, 2022, Vol. 48, No 2, pp. 126--137.}
 \newcommand{\dif}{\textrm{d}}

 \noindent
 \begin{tabular}{llllllllllllllllllllllllllllllllllllllllllllll}
 & & & & & & & & & & & & & & & & & & & & & & & & & & & & & & & & & & & & & &\\\hline\hline
 \end{tabular}

  \vskip 0.5cm
\centerline{\bf\large Estimation of the Parameters of the Spiral Pattern in the Galaxy}
\centerline{\bf\large Based on a Sample of Classical Cepheids}

   \bigskip
   \bigskip
  \centerline {V. V. Bobylev \footnote [1]{e-mail: vbobylev@gaoran.ru}}
   \bigskip

  \centerline{\small\it Pulkovo Astronomical Observatory, Russian Academy of Sciences,}

  \centerline{\small\it Pulkovskoe sh. 65, St. Petersburg, 196140 Russia}
 \bigskip
 \bigskip
 \bigskip

{\bf Abstract}---We consider a sample of Galactic classical Cepheids with highly accurate estimates of their distances taken from Skowron et al., where they were determined based on the period–luminosity relation. We have refined the geometric characteristics of two spiral arm segments—the Carina–Sagittarius and Outer ones. For this purpose, we have selected 269 Cepheids belonging to the Carina–Sagittarius arm with ages in the range 80--120 Myr. From them we have estimated the pitch angle of the spiral pattern $i=-11.9\pm0.2^\circ$ and the position of this arm $a_0=7.32\pm0.05$~kpc for the adopted $R_0=8.1\pm0.1$~kpc. In the Outer arm we have selected 343 Cepheids with ages in the range 120--300 Myr. From them we have
found $i=-11.5\pm0.5^\circ$ and $a_0=12.89\pm0.06$~kpc. Adhering to the model of a grand-design spiral pattern in the Galaxy with one pitch angle for all arms, we can conclude that this angle is close to $-12^\circ$.

 \bigskip
  DOI: 10.1134/S1063773722010017

 \subsection*{INTRODUCTION}
A huge number of publications are devoted to studying the spiral structure of the Galaxy using various objects and methods (see., e.g., Lin and Shu 1964; Lin et al. 1969; Y.M. Georgelin and
Y.P. Georgelin 1976; Taylor and Cordes 1993; Russeil
2003; Paladini et al. 2004; Popova and Loktin
2005; Dias and L\'epine 2005; Levine et al. 2006;
Moitinho et al. 2006; V\'azquez et al. 2008; Gerhard
2011; Efremov 2011; Bobylev and Bajkova
2014a; Hou and Han 2014, 2015; Dambis et al. 2015;
Reid et al. 2019; Xu et al. 2018a, 2021; Poggio
et al. 2021; Hao et al. 2021). Various stars, star
clusters and OB associations, hydrogen clouds, interstellar
dust clouds, masers, etc. are used as tracers
of the spiral structure. The methods based on both
the analysis of the spatial distribution of stars using their measured distances and the study of clumps of objects distributed along the Galactic equator are applied.

So far there is no universally accepted model of the Galaxy's grand-design spiral structure. Theorists commonly use the simplest model of a logarithmic spiral with two arms and a pitch angle of $5-7^\circ$. Highly accurate up-to-date data on the distribution of HI and HII clouds and masers with measured trigonometric parallaxes more likely suggest a four-armed model with a pitch angle of $10–14^\circ$. Ample convincing evidence for precisely the four-armed grand-design pattern was collected in the reviews by
Vall\'ee (1995, 2002, 2008, 2017).

Attempts, though unsuccessful, to estimate the variability of the pitch angle of the spiral pattern in the Galaxy were made (Hou and Han 2014). More complex models of the Galactic spiral structure were also proposed, for example, the model consisting of
a superposition of two-armed and four-armed structures near the Sun (L\'epine et al. 2001) or the model with two resonance circles R1 and R2 (Mel’nik and Rautiainen 2011; Mel’nik 2019).

Some authors prefer to analyze not the granddesign
structure, but separate spiral arm segments
close to the Sun with individual pitch angles (Nikiforov
and Veselova 2018; Veselova and Nikiforov
2020). The pitch angles estimated from the known
individual spiral arm segments lie within the range
$9-18^\circ$ (see, e.g., Bobylev and Bajkova 2014a; Griv
et al. 2017; Reid et al. 2019; Hao et al. 2021).

Two spiral arm segments close to the Sun, the Perseus and Carina–Sagittarius ones, which are
usually attributed to the grand-design structure, have been studied best. The Local arm located between the Perseus and Carina–Sagittarius arms
has also been well studied (Efremov 2011; Bobylev
and Bajkova 2014b; L\'epine et al. 2017; Vall\'ee 2018;
Hao et al. 2021; Xu 2021). The Local arm does not belong to the grand-design structure, but the Sun is currently virtually inside this local branch (spur) of the spiral structure.

The spiral pattern is assumed to rotate rigidly. The angular velocity of its rotation is variously estimated to be 15--30 km s$^{-1}$ kpc$^{-1}$ (Dias and L\'epine 2005;
Popova 2006; Gerhard 2011; Bobylev and Bajkova 2012). This means that the corotation radius
(the radius of the circle where the linear rotation
velocity of the spiral pattern coincides with the linear
rotation velocity of the Galaxy) is slightly farther from
the Sun toward the Galactic anticenter, somewhere
in the Perseus spiral arm or quite close to the Sun
(Barros et al. 2021).

A discussion of the present-day observed picture associated with the Galactic spiral structure can be found, for example, in the reviews by Hou and Han (2014), Vall\'ee (2017), Xu et al. (2018b), or Hou (2021).

The goal of this paper is to refine the parameters of the Galactic spiral pattern based on a large sample of classical Cepheids. Highly accurate distances to them were calculated based on the period–luminosity relation in Skowron et al. (2019) using mid-infrared photometry.

 \section*{METHOD}
The position of a star in a logarithmic spiral wave can be described by the following equation:
 \begin{equation}
 R=a_0 e^{(\theta-\theta_0)\tan i},
 \label{spiral-1}
 \end{equation}
where $R$ is the Galactocentric distance of the star, $\theta$ is the position angle of the star: $\tan\theta=y/(R_0-x)$, where $x, y$ are the heliocentric Galactic rectangular
coordinates of the star, with the $x$ axis being directed from the Sun to the Galactic center and the direction of the $y$ axis coincides with the direction of Galactic
rotation; $\theta_0$ is some arbitrarily chosen initial angle that we set equal to zero here, $\theta_0=0$; $a_0>0$ is the point of intersection of the $X$ axis directed away from
the Galactic center and passing through the Sun with the spiral, $i$ is the pitch angle of the spiral pattern ($i<0$ for a winding spiral) that is related to the remaining parameters as follows:
\begin{equation}
  \tan (|i|)=\frac{m\lambda}{2\pi R_0},
 \label{a-04}
\end{equation}
where $m$ is the number of spiral arms, $\lambda$ is the wavelength equal to the distance (in the radial direction) between the spiral arm segments in the solar neighborhood,
and $R_0$ is the Galactocentric distance of the Sun. In this paper we set $R_0$ equal to $8.1\pm0.1$~kpc, according to the review by Bobylev and Bajkova (2021a), where it was derived as a weighted mean of a large number of up-to-date individual estimates.

Given $a_0$ in Eq. (1), we can estimate the pitch angle $i:$
\begin{equation}
  \tan i=\frac{\ln (R/R_0)}{\theta}. 
 \label{spiral-04}
\end{equation}
For this purpose, we construct a position angle–distance logarithm diagram, where the spiral arm segments follow linear dependences. For each spiral arm segment we can estimate its $a_0$ and, consequently, $i.$ Note that in this method the estimate of the pitch angle $i$ does not depend on the number of spiral arms $m.$

\begin{figure}[t]
{ \begin{center}
   \includegraphics[width=0.7\textwidth]{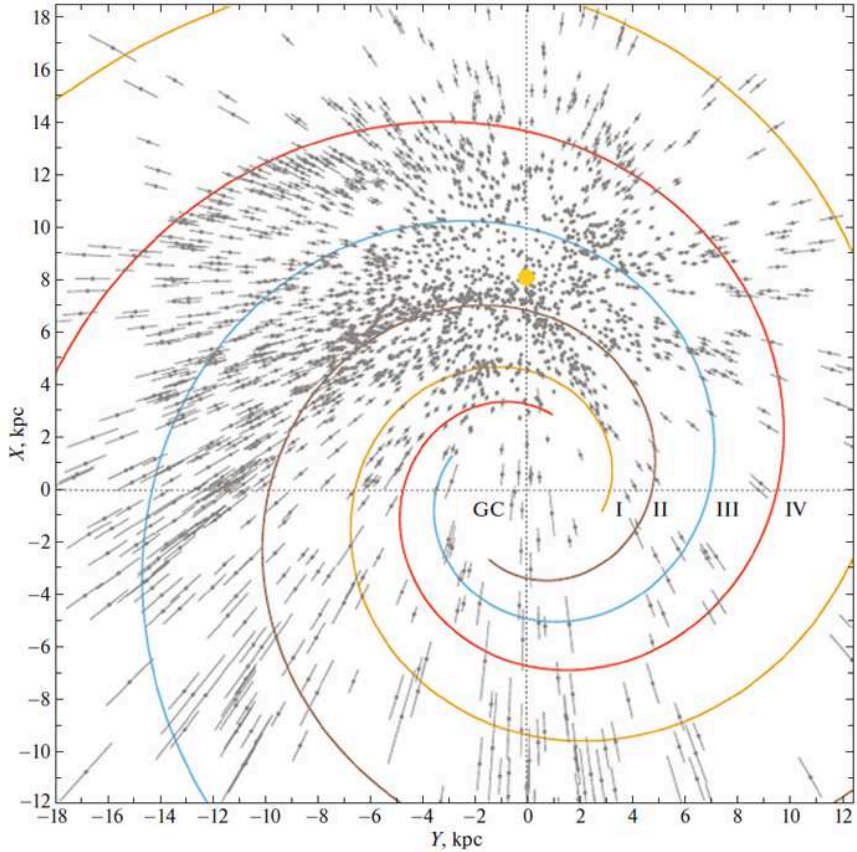}
  \caption{
Distribution of Cepheids on the Galactic $XY$ plane; the position of the Sun is marked by the yellow circle, GC denotes the Galactic center, the four-armed spiral pattern with the pitch angle $i=-13^\circ$ from Bobylev and Bajkova (2014a) is shown,
the spiral arms are numbered by Roman numerals.  
  }
 \label{f-1}
\end{center}}
\end{figure}
\begin{figure}[t]
{ \begin{center}
  \includegraphics[width=0.75\textwidth]{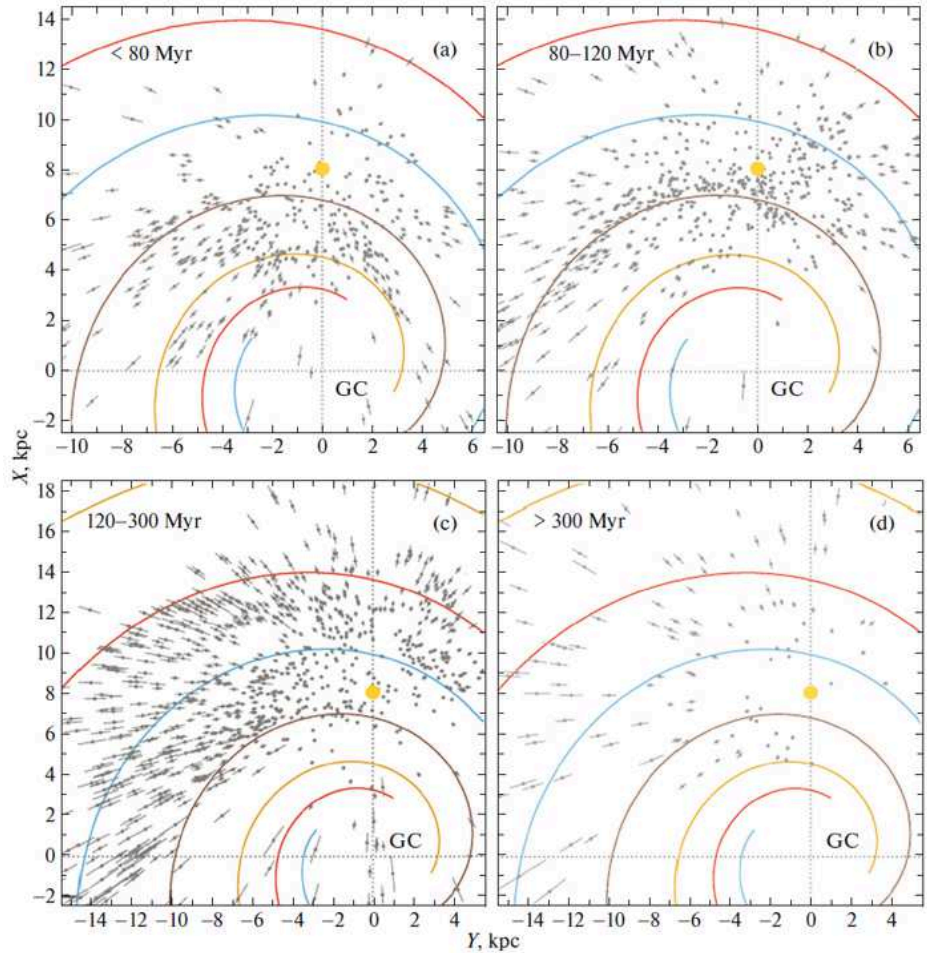}
  \caption{
Distribution of Cepheids younger than 80Myr (a), with ages in the range 80--120 Myr (b), with ages in the range 120--300 Myr (c), and older than 300 Myr (d) on the Galactic XY plane; the position of the Sun is marked by the yellow circle, GC denotes the Galactic center, the four-armed spiral pattern with the pitch angle $i=-13^\circ$ from Bobylev and Bajkova (2014a) is shown.
 }
 \label{f-2}
\end{center}}
\end{figure}

 \section*{DATA}
In this paper we use data on classical Cepheids from Skowron et al. (2019). These Cepheids were observed within the fourth stage of the OGLE (Optical Gravitational Lensing Experiment) program (Udalski et al. 2015). This catalogue contains distance, age,
pulsation period estimates and photometric data for
Cepheids. Their apparent magnitudes lie in the range $11^m<I<18^m$. Therefore, there is a slight deficit of bright and well-studied Cepheids known from earlier observations.

The heliocentric distances to 2214 Cepheids were calculated by Skowron et al. (2019) based on the period–luminosity relation. The specific relation was taken by them from Wang et al. (2018), in which it was refined from the light curves of Cepheids in
the mid-infrared, where the interstellar extinction is
much lower than that in the optical one. The Cepheid
ages in Skowron et al. (2019a) were estimated by the technique developed by Anderson et al. (2016), where the stellar rotation periods and metallicity indices were taken into account.

To solve the problem formulated here, we need a sample of stars the distances to which were determined using one calibration from homogeneous data.
Therefore, we do not add the data on other known
Cepheids the distances to which were determined
by other authors to the Cepheids from the list by
Skowron et al. (2019), especially since such other
Cepheids have been repeatedly used to determine the geometric characteristics of the Galactic spiral pattern (see, e.g., Dambis et al. 2015; Veselova and Nikiforov 2020).

Figure 1 presents the distribution of Cepheids in projection onto the Galactic $XY$ plane. The coordinate system in which the $X$ axis is directed to the Sun from the Galactic center and the direction of the $Y$ axis coincides with the direction of
Galactic rotation is used in the figure. The four-armed
spiral pattern with the pitch angle $i=-13^\circ$ (Bobylev and Bajkova 2014a) constructed with $R_0=8.1$ kpc is shown; the Roman numerals number the following spiral arm segments: Scutum (I), Carina--Sagittarius (II), Perseus (III), and the Outer arm (IV). We chose such a scale that some of the distant Cepheids remained outside the figure.

A concentration of a considerable number of stars to the Carina--Sagittarius spiral arm segment is clearly seen in Fig. 1. A concentration of Cepheids
in the third Galactic quadrant in the Outer arm can
also be noticed. The spiral pattern is plotted as an
example, and the objective of this paper is to refine
its parameter. It was shown in Bobylev and Bajkova
(2021b) that the distribution of Cepheids from
the catalogue by Skowron et al. (2019) in projection
onto the Galactic $XY$ plane differs greatly, depending
on the stellar age. Therefore, for the analysis it is
necessary to consider samples of Cepheids whose ages lie in narrow ranges. As can be seen from the figure, the Perseus spiral arm is represented very poorly in a wide solar neighborhood; the Local arm is also represented very poorly.

 \begin{table}[p]
 \caption[]{\small
Pitch angles $i$ and parameter $a_0$ for five spiral arm segments
 }
  \begin{center}  \label{t-i}  \small
  \begin{tabular}{|c|c|c|c|c|c|c|}\hline
 Parameter & I & II & III & IV & Local & Ref \\\hline

$i,$ deg &$-11.2\pm4.0$& $-9.3\pm2.2$&$-14.8\pm0.8$&$-11.5\pm1.9$&$-10.2\pm0.3$&(1)\\
$i,$ deg &$-10.2\pm1.0$&$-10.5\pm0.4$& $-7.9\pm1.2$&$-10.3\pm1.2$&             &(2)\\
$i,$ deg &$-13.1\pm2.0$& $-9.0\pm1.9$& $-9.5\pm2.0$& $-6.2\pm4.2$&$-11.4\pm1.9$&(3)\\
$i,$ deg &$-21.4\pm0.8$& $-9.9\pm1.3$&$-10.6\pm0.5$&$-18.6\pm6.9$&$-16.5\pm0.5$&(4)\\
$i,$ deg &             &$-16.3\pm1.0$& $-9.3\pm0.9$&             &$-10.8\pm0.9$&(5)\\
(20--40~Myr)         &&&&&& \\
$i,$ deg &             &$-16.2\pm1.0$& $-9.6\pm0.9$&             &$-10.5\pm0.9$&(5)\\
(60--80~Myr)         &&&&&& \\
$i,$ deg &  $-8~~~~~~$ &$-13.8~~~~~~$&$-11.9~~~~~~$&             & $-8.5~~~~~~$&(6)\\

$i,$ deg &$-18.7\pm0.8$& $-13.5\pm0.5$&$-9.0\pm0.1$&             &$-11.5\pm0.5$&(7)\\
$i,$ deg &$-15.1\pm0.7$& $-17.4\pm0.2$&$-7.0\pm0.3$&             &$-10.2\pm0.2$&(8)\\

$i,$ deg &$-11.7\pm0.9$& $-13.1\pm1.4$&$-6.2\pm1.6$& $-5.2\pm2.8$&$ -9.9\pm1.2$&(9)\\
 ($\sim$50~Myr)        &&&&&& \\
$i,$ deg &$ -8.9\pm2.1$& $ -8.3\pm0.8$&$-7.0\pm1.2$& $-9.7\pm1.1$&$ -7.0\pm0.6$&(9)\\
($\sim$100~Myr)         &&&&&& \\ \hline
$a_0,$~kpc       &$  4.5\pm0.2$& $ 6.8\pm0.3$&$  9.9\pm0.4$&$ 13.5\pm0.5$&$8.1\pm0.3$&(1)\\
($R_0=8$~kpc)    &&&&&& \\
$a_0,$~kpc       &$  5.2\pm0.1$& $ 6.4\pm0.1$&$  8.7\pm0.1$&$ 10.9\pm0.1$&           &(2)\\
($R_0=8$~kpc)    &&&&&& \\
$a_0,$~kpc       &$  4.9\pm0.1$& $ 6.0\pm0.1$&$  8.9\pm0.1$&$ 12.2\pm0.4$&$8.3\pm0.1$&(3)\\
($R_0=8.44$~kpc) &&&&&& \\
$a_0,$~kpc       &   6.5~~~~  &   6.9~~~~ &   9.6~~~~ &             & 8.5~~~~ &(6)\\
($R_0=8.15$~kpc) &&&&&& \\

$a_0,$~kpc       &$  5.9\pm0.1$& $ 7.2\pm0.1$&$ 10.6\pm0.1$&        &$8.3\pm0.1$&(7)\\
($R_0=8.35$~kpc) &&&&&& \\
$a_0,$~kpc       &$5.89\pm0.02$& $6.95\pm0.01$&$10.35\pm0.01$&     &$8.51\pm0.01$&(8)\\
($R_0=8.34$~kpc) &&&&&& \\

$a_0,$~kpc    &$6.07\pm0.04$&$6.78\pm0.05$&$9.74\pm0.09$&$12.02\pm0.17$&$8.19\pm0.05$&(9)\\
($R_0=8.08$~kpc) &&&&&& \\
 ($\sim$50~Myr) &&&&&& \\
$a_0,$~kpc    &$6.18\pm0.06$&$6.83\pm0.02$&$9.62\pm0.05$&$11.93\pm0.07$&$8.10\pm0.02$&(9)\\
($R_0=8.08$~kpc) &&&&&& \\
($\sim$100~Myr) &&&&&& \\
 \hline
 \end{tabular}\end{center}
 {\small
 (1) Bobylev and Bajkova (2014a); (2) Dambis et al. (2015); (3) Reid et al. (2019); (4) Nikiforov and Veselova (2018); (5) Hao et al. (2021); (6) Hou et al. (2021); (7) Xu et al. (2018b); (8) Zheng et al. (2019); (9) Veselova and Nikiforov (2020); I---the Scutum arm, II---the Carina--Sagittarius arm, III---the Perseus arm, IV---the Outer arm, and Local---the local arm (Orion arm).
 }
  \end{table}

Table 1 gives the pitch angles of five spiral arm segments and the parameter $a_0$ determined by various authors.

Bobylev and Bajkova (2014a), Reid et al. (2019), and Nikiforov and Veselova (2018) used data on masers with measured trigonometric parallaxes to
determine the parameters of the spiral structure. Only
to determine the parameters of the Outer arm did Bobylev and Bajkova (2014a) use data on several very young distant open star clusters together with masers.

Dambis et al. (2015) used a sample of classical Cepheids for this purpose. In this case, all calculations were performed in relative units $R/R_0.$ It was noted that the results remain valid at $R_0$ lying in the range 7.1--8 kpc. For the analysis these authors
used 565 bright classical Cepheids within 5~kpc of the
Sun. For this purpose, they took Cepheids from the catalogue by Berdnikov et al. (2000). The distances to the Cepheids were estimated with the period--luminosity relation using the procedure of Berdnikov et al. (1996) in the infrared K band in combination with the period--unreddened $(B-V )$ color from Dean et al. (1978). Dambis et al. (2015) estimated the distances to the Cepheids by a method independent of that from Skowron et al. (2019); brighter stars were used. They found the strong concentration of sampl stars toward the Carina--Sagittarius arm to be similar to our Fig.~1.

Xu et al. (2018b) gave an overview of the results obtained from various tracers of the spiral structure--- neutral and molecular hydrogen clouds, open star
clusters, masers with measured trigonometric parallaxes,
and O stars with trigonometric parallaxes from
the Gaia DR2 catalogue. Xu et al. (2018b) determined
the characteristics of the spiral structure from
a combination of data on the positions of 102 masers
and 583 O stars, where the masers and O stars were
taken with weights of 10 and unit weights, respectively

Zheng et al. (2019) determined the characteristics of the spiral structure from a combination of data on masers and OB2 stars, where the masers and
OB2 stars were taken with weights of 10 and unit
weights, respectively, just as in Xu et al. (2018b).
A huge sample of OB2 stars containing 14\,880
OB2 stars with parallaxes from the Gaia DR2 catalogue
was used here. Data on spectroscopically confirmed
candidates from VPHAS (VST Photometric $H_\alpha$ Survey, Drew et al. 2014) were used to select these stars from the Gaia DR2 catalogue. Because of such a large number of stars, the errors in $a_0$ turned out to be very small here.

Veselova and Nikiforov (2020) used data on 674 Cepheids from Mel’nik et al. (2015) to determine the parameters of several spiral arm segments.
This sample was divided into two parts in pulsation
period with a 4.6d boundary. The mean age is $\sim$50
and $\sim$100 Myr for relatively young and relatively
old Cepheids, respectively. Table~1 from Veselova
and Nikiforov (2020) gives the parameters for their
Carina--Sagittarius-1 and Outer-1 arms. We recalculated
the values of a0 taken from their Table 1 for $R_0=8.08$~kpc.

Hao et al. (2021) used data on 3794 open star clusters for which the mean distances were calculated from their trigonometric parallaxes from the
Gaia EDR3 catalogue. Table 1 gives two results of
these authors obtained from samples of clusters with
various ages.

Hou et al. (2021) jointly analyzed the data on giant molecular clouds, masers, HII regions, O stars, and young open star clusters. Note that Hou
et al. (2021) identified two independent segments
in arm II (Carina--Sagittarius arm) with slightly
differing characteristics. Table 1 gives the parameters
for this arm that we averaged over the two segments,
according to the data from Hou et al. (2021).

The estimates of the parameters presented in Table~1 were obtained mainly by analyzing the position angle--distance logarithm diagram. Of course, the
data sets differ. In this case, we can talk only about
satisfactory agreement between all estimates from
each of the spiral arm segments.

Figure 1 presented the distribution of all the
Cepheids considered in this paper. However, their
distribution on the $XY$ plane depends strongly on
the stellar age. This can be clearly seen from Fig.~2,
where the distribution of four samples of Cepheids
with various ages in projection onto the Galactic $XY$
plane is presented.

An analysis of the distributions of Cepheids in
Fig. 2 leads to a number of important conclusions.
The Cepheids older than 300 Myr (Fig. 2d) are of no
interest for refining the geometric characteristics of
the spiral pattern. No concentration toward the spiral
arms is seen in their distribution. There are a total of
126 Cepheids in this sample.

A structure elongated from the Sun along the line
of sight at an angle of about 0--40$^\circ$ is clearly seen
in the distribution of Cepheids younger than 80 Myr
(Fig. 2a). These Cepheids were apparently observed
in the transparency window. Using these stars
can lead to a significant distortion of the geometric
parameters of the Carina--Sagittarius spiral arm
(arm II). Using such stars in Hou et al. (2021) forced
these authors to consider two Carina--Sagittarius
arm segments with differing characteristics. There
are a total of 470 Cepheids younger than 80 Myr; in
this paper we decided not to use these stars.

A concentration of Cepheids toward the Carina--Sagittarius arm can be seen in both Fig. 2b and Fig. 2c. Therefore, we can analyze these samples
both separately and jointly to refine the geometric
characteristics of the Carina--Sagittarius spiral arm.
In our opinion, it is more advantageous to analyze the
samples separately, despite the partial loss of stars,
because, in this case, we have more homogeneous
samples with regard to the stellar age.

A concentration of Cepheids toward the Outer
spiral arm can be seen only in Fig. 2c. There is a pronounced
gap---an empty space between the Perseus
and Outer arms in the third Galactic quadrant. It
can be seen that the Cepheids are not located at the
center of the Outer arm shown in the figure and found
by analyzing masers (Bobylev and Bajkova 2014a).
There are quite a few such Cepheids in the sample
with ages of 120--300 Myr located in the figure along
the inner edge of the Outer arm. Using them can lead
to a significant refinement of the characteristics of this
spiral arm.

It is impossible to reliably identify a sufficient number of Cepheids belonging to the Perseus and Scutum spiral arm segments without additional information about the stellar line-of-sight velocities. Therefore, in this paper we attempt to refine the
geometric characteristics of two spiral arms~--- the Carina–Sagittarius and Outer ones.

\begin{figure}[t]
{ \begin{center}
  \includegraphics[width=0.75\textwidth]{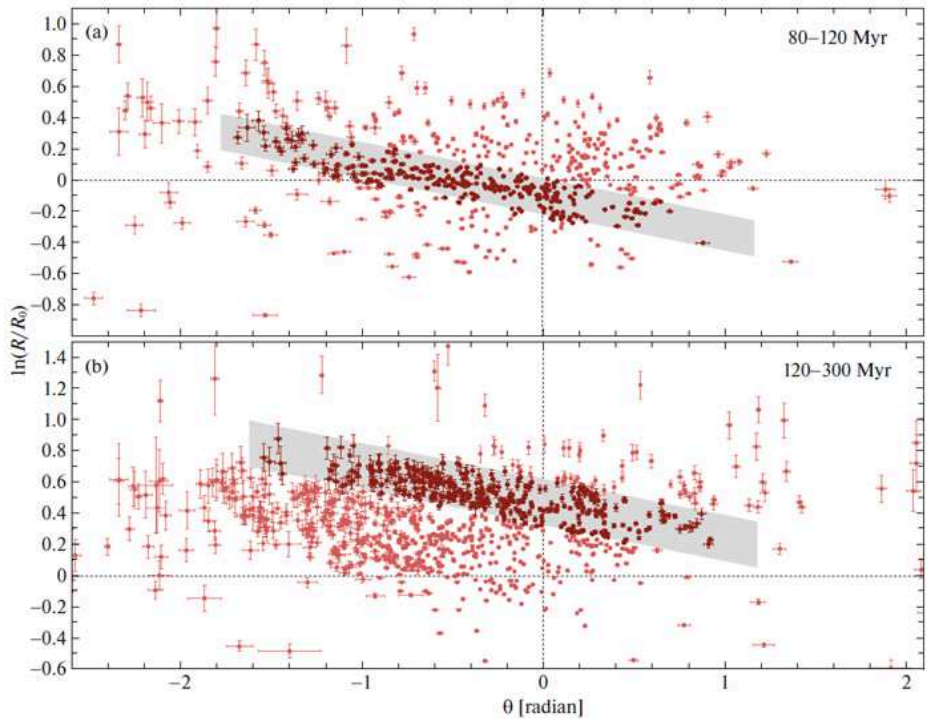}
  \caption{
Position angle–distance logarithm diagram for the Cepheids with ages in the ranges 80--120 (a) and 120--300 (b)Myr; the shading on each graph indicates the star preselection region.}
 \label{f-lnR}
\end{center}}
\end{figure}

 \section*{RESULTS AND DISCUSSION}
Figure 3 presents the position angle--distance logarithm diagrams for two samples of Cepheids. The boundaries for the preselection of Cepheids with ages in the range 80--120 Myr, where the mean age is 100~Myr, and in the range 120--300~Myr with a mean
age of 200~Myr are shown.

The following constraints were used for the selection of stars in the Carina--Sagittarius arm: $-1.7<\theta<1.1$~rad, while $\ln(R/R_0)$ was bounded by two lines from above and from below: $\theta\cdot\tan(-13^\circ)+0.02<\ln(R/R_0)<\theta\cdot\tan(-13^\circ)-0.25$.

For the selection of stars in the Outer arm we took the angle $\theta$ from the range $-1.6<\theta<1.1$~rad and used the following constraints for $\ln(R/R_0)$:
$\theta\cdot\tan(-13^\circ)+0.6208<\ln(R/R_0)<\theta\cdot\tan(-13^\circ)+0.3120$.
Compared to the estimate of a0 found for
the Outer arm in Bobylev and Bajkova (2014a) (see
Table 1), here we used a lower value of this quantity,
i.e., the selection zone is shifted closer to the Sun.

The width of the zone corresponding to the constraints on $\ln(R/R_0)$ was taken with a large margin--- it exceeds the known diameter of the arms by a factor
of 5--6. For example, having analyzed the distribution
of a relatively small number of masers with measured
trigonometric parallaxes, Reid et al. (2019) estimated
the widths of the Carina--Sagittarius and Outer arms
to be 0.27 and 0.65 kpc, respectively. As can be seen
from Fig. 3a, the bulk of the Cepheids stretches in
the selection zone for the Carina–Sagittarius arm
in the form of a thin string and there is much space
with a relatively low star density. Thus, the width
of this string in the Carina--Sagittarius arm is close
to the estimate of Reid et al. (2019). At the same
time, the width of the selection region chosen by us
here exceeds the estimate of Reid et al. (2019) by a
factor of 3--4. The Cepheids of the Outer arm have
a considerably older age ($\sim$200 Myr) and have had
time to recede far from their birthplaces. As can be
seen from Fig. 3b, the selected stars fill the selection
zone almost uniformly. The width of the selection
region here is about 4 kpc, which also exceeds
considerably the estimate of Reid et al. (2019).

Table 2 gives the parameters $a$ and $b$ found through the least-squares solution of the system of conditional equations $\ln(R/R_0)=a\cdot\theta+b.$ The solutions were
found both with unit weights and with weights $w_i=1/\sigma^2_{\ln(R/R_0)}$, where $i=1,...,n_\star$, and $n_\star$ is the number of stars used in the solution. The table also
gives the values of the parameter $a_0$ entering into the spiral equation (1) and marking the Galactocentric position of the arm center on the $X$ axis.

As can be seen from Table 2, both values of the
pitch angle $i$ are in excellent agreement between each
other. There is also agreement with the results presented
in Table 1 within the limits of the random errors declared there. Note that the pitch angle of the Outer
arm found in this paper was determined very reliably,
because its error is small.

  \begin{table*}
   \begin{center}
    \caption{
The parameters of the linear dependence $\ln(R/R_0)=a\cdot\theta+b$ found with unit weights and weights $w_i=1/\sigma^2_{\ln(R/R_0)}$ are given in the upper and lower parts of the table, respectively, $n_\star$ is the number of Cepheids used
    }
   \label{t2}
   {\small
   \begin{tabular}{|l|c|c|c|r|r|}      \hline
 Arm          & $n_\star$ & $a$ & $b$ & $i,$ deg & $a_0$, kpc\\\hline
 II (Carina--Sagittarius)&269&$-0.212\pm0.008$& $-0.104\pm0.005$& $-11.98\pm0.45$&$ 7.29\pm0.05$\\
 IV  (Outer)     &343&$-0.206\pm0.009$& $+0.458\pm0.005$& $-11.66\pm0.49$&$12.81\pm0.06$\\
 \hline
 II (Carina--Sagittarius)&269&$-0.211\pm0.004$& $-0.102\pm0.002$& $-11.94\pm0.22$&$ 7.32\pm0.05$\\
 IV  (Outer)     &343&$-0.203\pm0.008$& $+0.465\pm0.005$& $-11.46\pm0.47$&$12.89\pm0.06$\\
 \hline
      \end{tabular}}
     \end{center}
   \end{table*}
\begin{figure}[t]
{ \begin{center}
  \includegraphics[width=0.65\textwidth]{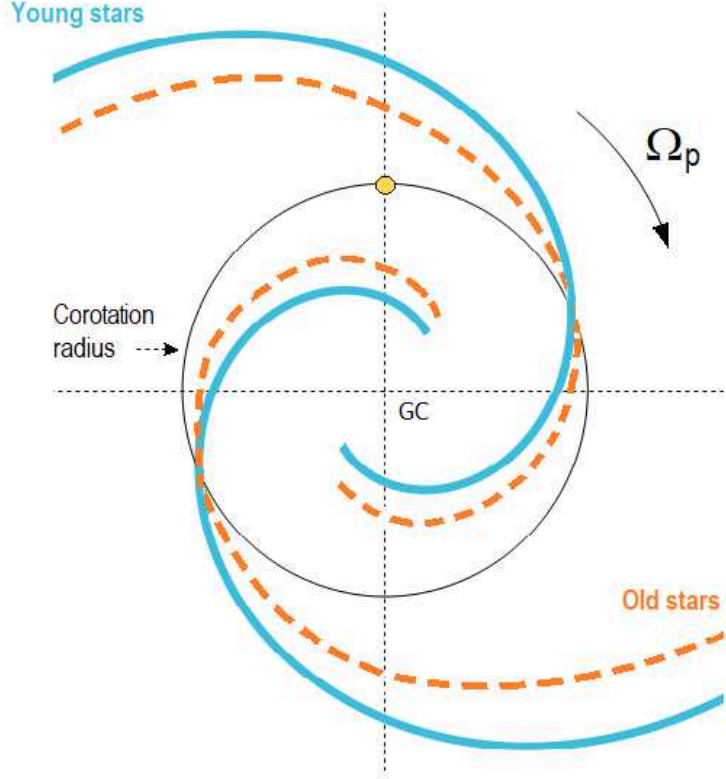}
  \caption{
Scheme of the distribution of young and old stars relative to the corotation region; the corotation circle is indicated by the thin solid line on which the yellow circle indicates the observer’s location; the place of current star formation is marked by the blue thick line; the distribution of older stars is indicated by the orange dashed line. The direction of rotation of the spiral pattern with an angular velocity $\Omega_p$ is shown, GC is the Galactic center.}
 \label{f-4}
\end{center}}
\end{figure}
\begin{figure}[t]
{ \begin{center}
   \includegraphics[width=0.7\textwidth]{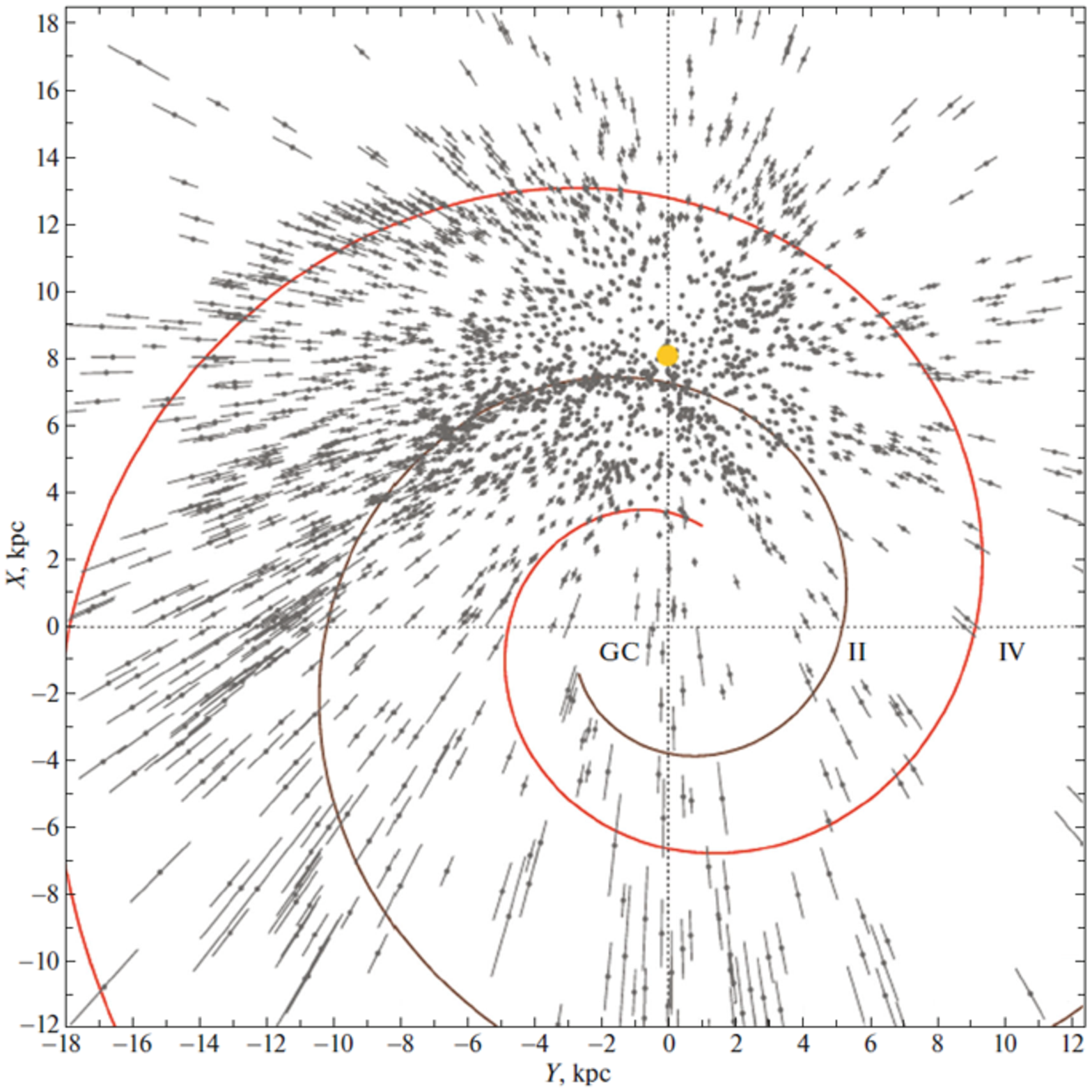}
  \caption{
Distribution of Cepheids on the Galactic $XY$ plane, the position of the Sun is marked by the yellow circle, GC is the Galactic center, the two spiral arms constructed with the parameters found (the upper part of Table 2) are shown.
 }
 \label{f-5}
\end{center}}
\end{figure}

Adhering to the model of a grand-design spiral pattern in the Galaxy with one pitch angle for all arms, we can conclude that this angle is close to $-12^\circ$. In future, it is desirable, of course, to determine this angle from a larger number of observed spiral
arm segments. Therefore, here we do not calculate the mean value of this angle from the results of our analysis of only two spiral arm segments.

The same can also be said about the agreement
between the determinations of a0. It can be noted
that, in our case, the Carina--Sagittarius arm (II) is
closer to the Sun, while the Outer arm (IV) is farther
compared to the results of the analysis of other samples
of Cepheids (Dambis et al. 2015). In our opinion,
this means that the point is not in the distance scales,
but in the distribution of Cepheids with various ages.

Figure 4 was prepared for a more detailed explanation
of this thesis. It presents a scheme of the distribution
of young and old stars in the spiral arms in
various zones relative to the corotation region. This is
a well-known scheme (Martinez-Garcia et al. 2009;
Hou and Han 2015; Xu et al. 2018b) reflecting the
predictions of the quasi-stationary case of a spiral
density wave (Roberts 1969). It was adapted by us
for a location closer to that of the Outer arm.

Bobylev and Bajkova (2014a) determined the parameters
of the spiral pattern from masers located
in active star-forming regions. Therefore, the spiral
pattern indicated in Figs. 1 and 2 marks the place of
current star formation. The Cepheids with ages from
the range 80--120 Myr are considerably older than
these masers, whose age is only a few Myr. Therefore,
we have excellent agreement between the positions of
the Cepheids with ages fromthe range 80--120 Myr in
Fig. 2b and on the scheme of Fig. 4 inside the corotation
circle. Thus, the value of $a_0=7.29\pm0.05$~kpc found from them exceeds $a_0=6.8\pm0.3$~kpc found from masers in Bobylev and Bajkova (2014a).

The Cepheids with ages from the range 120--300 Myr, of course, have had time to recede far from their birthplaces. Despite this, here there is also
qualitative agreement between their positions relative
to the Outer arm in Fig. 1, Fig. 2c, and on the scheme
of Fig. 4 outside the corotation circle.

It can be concluded that the pitch angles $i$ (Table~2) were determined excellently. In contrast, the values of $a_0$ in Table~2 characterize not the positions of
the spiral arm centers, but the positions of the regions
where the Cepheids were formed.

To make sure that the results are reproducible, we calculated the parameters of the spiral pattern with various weights and for wider constraints on the
selection band width. A new selection was made
already with a selection band pitch angle of $-12^\circ$,
whose value was found in this paper earlier.

Apart fromunit weights, we chose weights in three
more forms: $w_i=1/\sigma^2_r$, $w_i=\sigma^2_0/(\sigma^2_0+\sigma^2_r)$, where
$\sigma_0=0.01$~kpc, and $w_i=1/\sigma^2_{\ln(R/R_0)}$. Since we apply
the weights, the selection band of Cepheids from the Carina--Sagittarius arm was expanded in position angle $-2.5<\theta<1.1$~rad to include stars with large distance errors from the third and partly fourth quadrants (Fig. 2b).

The results are presented in Table~3 for the Carina--Sagittarius arm and in Table~4 for the Outer arm. The selection band width $\Delta R$ in kpc is given
in these tables instead of the boundary conditions
expressed in logarithms and occupying a large width
in the table’s column. It follows from these tables that
the pitch angle in both cases is within the range 11--12$^\circ$. This angle is determined particularly reliably from the Cepheids of the Outer arm with various weights in the upper part of Table~3. Expanding the selection region increases the dispersion of the
results, particularly the pitch angles.

Figure 5 presents the distribution of the entire sample of Cepheids on the $XY$ plane, where the two spiral arms constructed with the parameters found (the upper part of Table 2) are shown. It can be seen that both arms pass excellently through the data in
the second and third Galactic quadrants. In this case,
all of what has been said above when discussing the
scheme of Fig.~4 should be taken into account.

  \begin{table*}
   \begin{center}
    \caption{
The parameters of the dependence $\ln(R/R_0)=a\cdot\theta+b$ found at various widths $\Delta R$ of the Cepheid selection band in the Carina--Sagittarius arm at a band pitch angle of $-12^\circ$
    }
   \label{t3}
   {\small
   \begin{tabular}{|l|c|c|c|c|c|c|}      \hline
 $w_i,$  & $n_\star$ & $a$ & $b$ & $i,$ deg & $\Delta R$, kpc\\\hline
 $w_i=1$
    &282&$-0.205\pm0.007$& $-0.105\pm0.005$& $-11.61\pm0.39$& 1.95\\
 $w_i=1/\sigma^2_r$
    &282&$-0.188\pm0.008$& $-0.099\pm0.005$& $-10.65\pm0.48$& 1.95\\
 $w_i=\sigma^2_0/(\sigma^2_0+\sigma^2_r)$
    &282&$-0.185\pm0.009$& $-0.099\pm0.004$& $-10.48\pm0.52$& 1.95\\

 $w_i=1/\sigma^2_{\ln(R/R_0)}$
    &282&$-0.207\pm0.004$& $-0.102\pm0.003$& $-11.69\pm0.21$& 1.95\\
 \hline
 $w_i=1$
    &302&$-0.203\pm0.007$& $-0.102\pm0.005$& $-11.46\pm0.41$& 2.20\\
 $w_i=1/\sigma^2_r$
    &302&$-0.181\pm0.009$& $-0.095\pm0.005$& $-10.28\pm0.50$& 2.20\\

 $w_i=\sigma^2_0/(\sigma^2_0+\sigma^2_r)$
    &302&$-0.177\pm0.010$& $-0.095\pm0.005$& $-10.05\pm0.54$& 2.20\\

 $w_i=1/\sigma^2_{\ln(R/R_0)}$
    &302&$-0.205\pm0.004$& $-0.099\pm0.003$& $-11.57\pm0.22$& 2.20\\
 \hline
      \end{tabular}}
     \end{center}
   \end{table*}
  \begin{table*}
   \begin{center}
    \caption{
The parameters of the dependence $\ln(R/R_0)=a\cdot\theta+b$ found at various widths $\Delta R$ of the Cepheid selection band in the Outer arm at a band pitch angle of $-12^\circ$
    }
   \label{t4}
   {\small
   \begin{tabular}{|l|c|c|c|c|c|c|}      \hline
 $w_i,$  & $n_\star$ & $a$ & $b$ & $i,$ deg & $\Delta R$, kpc\\\hline
 $w_i=1$
    &343&$-0.206\pm0.009$& $0.458\pm0.005$& $-11.66\pm0.49$& 4.12\\
 $w_i=1/\sigma^2_r$
    &343&$-0.222\pm0.011$& $0.413\pm0.004$& $-12.52\pm0.64$& 4.12\\
 $w_i=\sigma^2_0/(\sigma^2_0+\sigma^2_r)$
    &343&$-0.212\pm0.011$& $0.425\pm0.004$& $-11.97\pm0.62$& 4.12\\
 $w_i=1/\sigma^2_{\ln(R/R_0)}$
    &343&$-0.203\pm0.008$& $0.465\pm0.005$& $-11.46\pm0.47$& 4.12\\
 \hline
 $w_i=1$
    &372&$-0.192\pm0.009$& $0.457\pm0.005$& $-10.89\pm0.52$& 4.63\\
 $w_i=1/\sigma^2_r$
    &372&$-0.216\pm0.012$& $0.402\pm0.004$& $-12.18\pm0.67$& 4.63\\
 $w_i=\sigma^2_0/(\sigma^2_0+\sigma^2_r)$
    &372&$-0.203\pm0.011$& $0.417\pm0.005$& $-11.48\pm0.64$& 4.63\\
 $w_i=1/\sigma^2_{\ln(R/R_0)}$
    &372&$-0.187\pm0.009$& $0.466\pm0.005$& $-10.59\pm0.49$& 4.63\\
 \hline
 $w_i=1$
    &415&$-0.181\pm0.010$& $0.452\pm0.006$& $-10.26\pm0.55$& 5.42\\
 $w_i=1/\sigma^2_r$
    &415&$-0.237\pm0.013$& $0.378\pm0.005$& $-13.34\pm0.73$& 5.42\\
 $w_i=\sigma^2_0/(\sigma^2_0+\sigma^2_r)$
    &415&$-0.212\pm0.012$& $0.401\pm0.005$& $-11.98\pm0.69$& 5.42\\
 $w_i=1/\sigma^2_{\ln(R/R_0)}$
    &415&$-0.171\pm0.009$& $0.464\pm0.006$& $- 9.71\pm0.52$& 5.42\\
 \hline
      \end{tabular}}
     \end{center}
   \end{table*}

The parameters of the Outer arm are of great importance,
because they have been obtained for the first
time with a high accuracy from a large sample of stars
with reliable distance estimates. Indeed, Bobylev and
Bajkova (2014a) analyzed the Outer arm using three
masers in combination with data on 12 young star
clusters. Reid et al. (2019) used 11 masers in this
arm. Nikiforov and Veselova (2018) estimated the
parameters of theOuter arm from six masers. Dambis
et al. (2015) selected 99 Cepheids in the Outer arm for
their analysis.

Veselova and Nikiforov (2020) used 10 young and
89 old Cepheids to determine the parameters of the
Outer arm. Note that the pitch angle of the Outer
arm found here agrees best precisely with the estimates
of Dambis et al. (2015) and Veselova and
Nikiforov (2020) obtained by them from a sample of
old Cepheids.

 \section*{CONCLUSIONS}
We considered a sample of Galactic classical Cepheids with highly accurate estimates of their distances from Skowron et al. (2019), which were calculated by these authors based on the period--luminosity relation.

To refine the geometric characteristics of the Carina--Sagittarius and Outer arm segments, we used Cepheids with ages in the range 80--300 Myr.
For this purpose, we selected 269 Cepheids belonging
to the Carina--Sagittarius arm with ages in the
range 80--120 Myr. From them we estimated the
pitch angle of the spiral pattern $i=-12.0\pm0.5^\circ$
and the position of the center of this arm on the
Galactic center--Sun axis $a_0=7.29\pm0.05$~kpc for
the adopted $R_0=8.1\pm0.1$~kpc with the application
of unit weights. Based on 343 Cepheids in the
Outer arm with ages of 120--300 Myr, we found
$i=-11.7\pm0.5^\circ$ and $a_0=12.81\pm0.06$~kpc.

Our calculations were repeated with the application of weights $w_i=1/\sigma^2_{\ln(R/R_0)}$. In this case, we obtained the following estimates: $i=-11.9\pm0.2^\circ$ and $a_0=7.32\pm0.05$~kpc for the Carina--Sagittarius arm and $i=-11.5\pm0.5^\circ$ and $a_0=12.89\pm0.06$~kpc
for the Outer arm.

The parameters of the Outer arm are of greatest
interest, because they have been obtained for the first
time based on such a large and homogeneous sample
of stars with reliable distance estimates.

The stability of the spiral pattern parameters found
was confirmed by the calculations performed with
various weights under various wider constraints on
the Cepheid selection bands in the arms being analyzed.

Adhering to the model of a grand--design spiral pattern in the Galaxy with one pitch angle for all arms, we can conclude that this angle is close to $-12^\circ$. In future, it is desirable, of course, to determine this angle based on a larger number of observed spiral arm segments.

\bigskip{ACKNOWLEDGMENTS}

I am grateful to the referee for the useful remarks
that contributed to an improvement of the paper.

\bigskip\medskip{REFERENCES}{\small

1. R. I. Anderson, H. Saio, S. Ekstr\"om, C. Georgy, and
G. Meynet, Astron. Astrophys. 591, A8 (2016).

2. D. A. Barros, A. Perez-Villegas, T. A. Michtchenko,
and J. R. D. Lepine, Front. Astron. Space Sci. 8, 48 (2021).

3. L. N. Berdnikov, Odessa Astron. Publ. 18, 23 (2006).

4. L. N. Berdnikov, O. V. Vozyakova, and A. K. Dambis, Astron. Lett. 22, 839 (1996).

5. V. V. Bobylev and A. T. Bajkova, Astron. Lett. 38, 638 (2012).

6. V. V. Bobylev and A. T. Bajkova, Mon. Not. R. Astron. Soc. 437, 1549 (2014a).

7. V. V. Bobylev and A. T. Bajkova, Astron. Lett. 40, 783 (2014b).

8. V. V. Bobylev and A. T. Bajkova, Astron. Rep. 65, 498 (2021a).

9. V. V. Bobylev and A. T. Bajkova, Astron. Lett. 47, 534 (2021b).

10. A. K. Dambis, L. N. Berdnikov, Yu. N. Efremov,
et al., Astron. Lett. 41, 489 (2015).

11. J. F.Dean, P. R. Warren, and A. W. J. Cousins, Mon.
Not. R. Astron. Soc. 183, 569 (1978).

12. W. S.Dias and J. R.D. L\'epine, Astrophys. J. 629, 825 (2005).

13. J. E. Drew, E. Gonzalez-Solares, R. Greimel, 
et al., Mon. Not. R. Astron. Soc. 440, 2036 (2014).

14. Yu. N. Efremov, Astron.Rep. 55, 105 (2011).

15. Y. M. Georgelin and Y. P. Georgelin, Astron. Astrophys. 49, 57 (1976).

16. O. Gerhard, Mem. Soc. Astron. Ital. Suppl. 18, 185 (2011).

17. E. Griv, I.-G. Jiang, and L.-G. Hou, Astrophys. J. 844, 118 (2017).

18. C. J. Hao, Y. Xu, L. G. Hou, 
 arXiv astroph: 2107.06478 (2021).

19. L. G. Hou and J. L. Han, Astron. Astrophys. 569, 125 (2014).

20. L. G. Hou and J. L. Han, Mon. Not. R. Astron. Soc. 454, 626 (2015).

21. L. G. Hou, Front. Astron. Space Sci. 8, 103 (2021).

22. J. R. D. L\'epine, Yu. N. Mishurov, and S. Yu. Dedikov,
Astrophys. J. 546, 234 (2017).

23. J. R. D. L\'epine, T. A.Michtchenko, D. A. Barros, and
R. S. S. Vieira, Astrophys. J. 843, 48 (2017).

24. E. S. Levine, L. Blitz, and C. Heiles, Science (Washington, DC, U. S.) 312, 1773 (2006).

25. C. C. Lin and F. H. Shu, Astrophys. J. 140, 646 (1964).

26. C. C. Lin, C. Yuan, and F. H. Shu, Astrophys. J. 155, 721 (1969).

27. E. E. Martinez-Garcia, R. A. Gonz\'alez-L\'opezlira, and G.-A. Bruzual, 
 Astrophys. J. 694, 512 (2009).

28. A. M. Mel’nik and P. Rautiainen, Mon. Not. R. Astron. Soc. 418, 2508 (2011).

29. A. M. Mel’nik, P. Rautiainen, L. N. Berdnikov, 
 et al., Astron. Nachr. 336, 70 (2015).

30. A. M. Melnik, Mon. Not. R. Astron. Soc. 485, 2106 (2019).

31. A. Moitinho, R. A. V\'azquez, G. Carraro, 
 et al.. Mon. Not. R. Astron. Soc. 368, L77 (2006).

32. I. I. Nikiforov and A. V. Veselova, Astron. Lett. 44, 81 (2018).

33. R. Paladini, R. Davies, and G. DeZotti, Mon. Not. R. Astron. Soc. 347, 237(2004).

34. E. Poggio, R. Drimme, T. Cantat-Gaudin, P. Ramos,
et al., Astron. Astrophys. 651, 104 (2021).

35. M. E. Popova and A. V. Loktin, Astron. Lett. 31, 171 (2005).

36. M. E. Popova, Astron. Lett. 32, 244 (2006).

37. M. J. Reid, K. M. Menten, A. Brunthaler,
et al., Astrophys. J. 885, 131 (2019).

38. W. W. Roberts, Astrophys. J. 158, 123 (1969).

39. D. Russeil, Astron. Astrophys. 397, 133 (2003).

40. D. M. Skowron, J. Skowron, P. Mr\'oz, A. Udalski, et al., Science 
 (Washington, DC, U. S.) 365, 478 (2019).

41. J. H. Taylo, and J. M. Cordes, Astrophys. J. 411, 674 (1993).

42. A. Udalski, M. K. Szyma\'nski, and G. Szyma\'nski, Acta Astron. 65, 1 (2015).

3. J. P. Vall\'ee, Astrophys. J. 454, 119 (1995).

44. J. P. Vall\'ee, Astrophys. J. 566, 261 (2002).

45. J. P. Vall\'ee, Astron. J. 135, 1301 (2008).

46. J. P. Vall\'ee, New Astron. Rev. 79, 49 (2017).

47. J. P. Vall\'ee, Astrophys. Space Sci. 363, 243 (2018).

48. R. A. V\'azquez, J. May, G. Carraro, L. Bronfman, A. Moitinho, and G. Baume, Astrophys. J. 672, 930 (2008).

49. A. V. Veselova and I. I. Nikiforov, Res. Astron. Astrophys. 20, 209 (2020).

50. S. Wang, X. Chen, R. de Grijs, and L. Deng, Astrophys. J. 852, 78 (2018).

51. Y. Xu, S. Bian, M. J. Reid, 
et al., Astron. Astrophys. 616, L15 (2018a).

52. Y. Xu, L.-G. Hou, and Y.-W. Wu, Res. Astron. Astrophys. 18, 146 (2018b).

53. Y. Xu, L. G.Hou, S. Bian, 
 et al., Astron. Astrophys. 645, L8 (2021).

 }
\end{document}